\documentclass[a4paper]{article} 

\usepackage[utf8]{inputenc}
\usepackage[T1]{fontenc}
\usepackage{graphicx, wrapfig, floatflt, subfig}%
\usepackage{mathptmx}      
\usepackage[para]{footmisc} 
\usepackage{rotating}
\usepackage{stfloats}
\usepackage{enumerate}
\usepackage[most]{tcolorbox}
\usepackage[english]{babel}
\usepackage{nth}
\usepackage{natbib}
\usepackage{soul}
\usepackage[strings]{underscore}
\newcommand{\source}[2]{\raggedleft{}\vspace*{-5mm}\caption*{ \textmd{\scriptsize{Data source: {#1}.\hfill Tool:{#2}}}}}
\def\BibTeX{{\rm B\kern-.05em{\sc i\kern-.025em b}\kern-.08emT\kern-.1667em\lower.7ex\hbox{E}\kern-.125emX}}

\newtcolorbox[blend into=figures]{card}[2][]{enhanced,
float=tbp,title={#2},
colframe=gray!75!black, colback=yellow!5!white,#1
}
\providecommand{\keywords}[1]
{
  \small	
  \textbf{\textit{Keywords---}} #1
}
\begin{document}

\title{Research Frontiers in Transfer Learning\footnote{pre-print.}\\a systematic and bibliometric review}

\author{Frederico Guth         \and
  Teófilo Emidio de Campos 
}

\maketitle

\begin{abstract}
Humans can learn from very few samples, demonstrating an outstanding generalization ability that learning algorithms are still far from reaching. Currently, the most successful models demand enormous amounts of well-labeled data, which are expensive and difficult to obtain, becoming one of the biggest obstacles to the use of machine learning in practice. This scenario shows the massive potential for Transfer Learning, which aims to harness previously acquired knowledge to the learning of new tasks more effectively and efficiently. In this systematic review, we apply a quantitative method to select the main contributions to the field and make use of bibliographic coupling metrics to identify research frontiers. We further analyze the linguistic variation between the ``classics'' of the field and the ``frontier'' and map promising research directions.

\end{abstract}

\section{Introduction}
\label{intro}
Currently, machine learning algorithms can recognize objects, people, and places at a super-human accuracy~\citep{feifei}; they can diagnose skin cancer better than dermatologists~\citep{guth2018skin}, and can even see through walls using radio signal analysis~\citep{wifi}. Despite all that, most successful models demand astronomical amounts of well-labeled data that are expensive and difficult to gather. That is because the standard model training procedure starts \emph{tabula rasa}, i.e., with random initialization of model parameters~\citep{Ruder2019Neural}.

Learning this way, from a blank state, is contrary to the way humans do. Every day, we transfer knowledge: knowing how to play the piano makes it easier to learn pipe organ; knowing Portuguese makes it easier to learn Spanish. People use previously obtained knowledge to more effectively and efficiently learn new things \citep{PanYang}. Learning algorithms are still far from reaching this outstanding generalization capability. Recent studies~\citep{DBLP:journals/corr/JiaL17} show that current algorithms hardly generalize further than the data seen during training.

\begin{figure}
\centering
\subfloat[]{{\includegraphics[width=.4\textwidth]{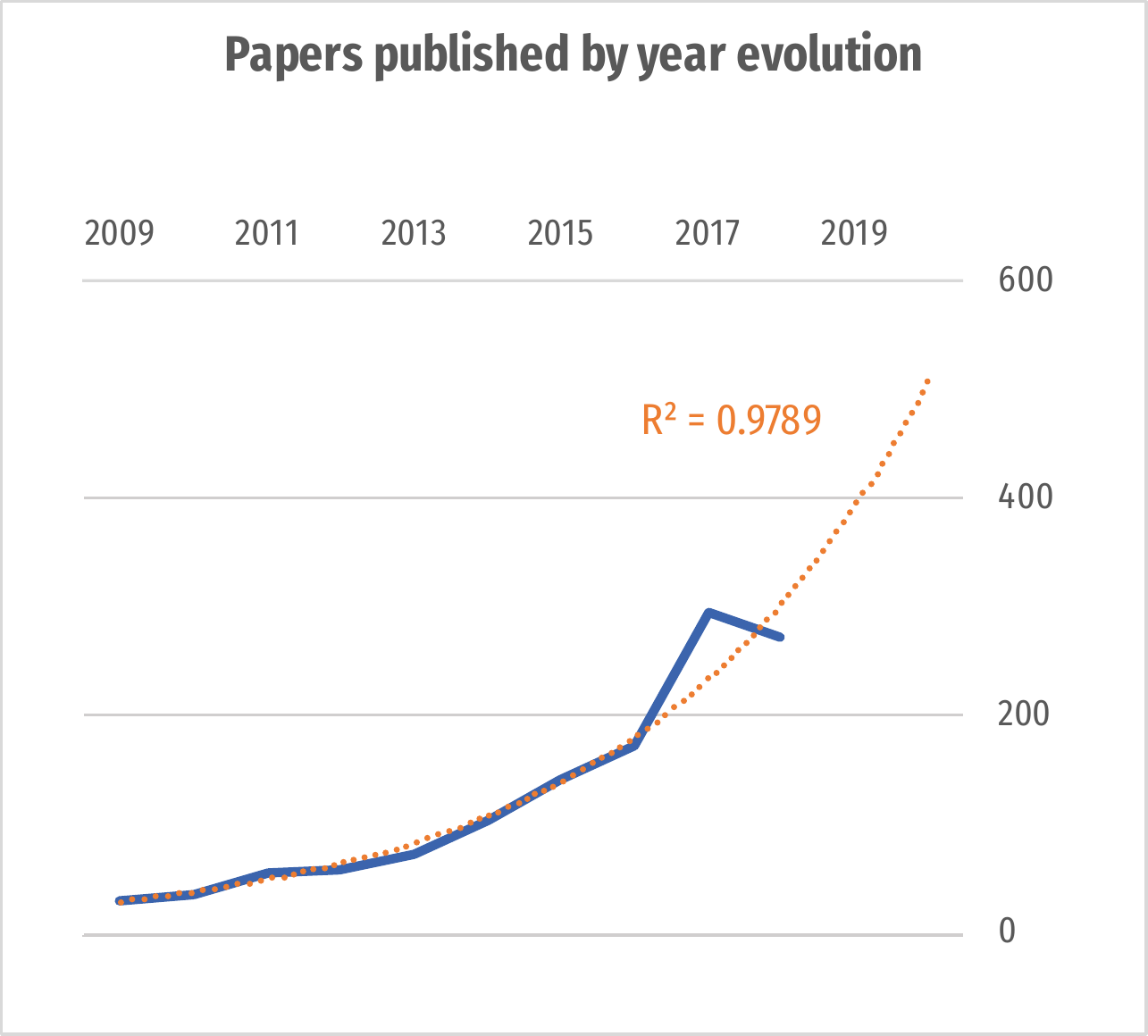} }}%
\qquad
\subfloat[]{{\includegraphics[width=.4\textwidth]{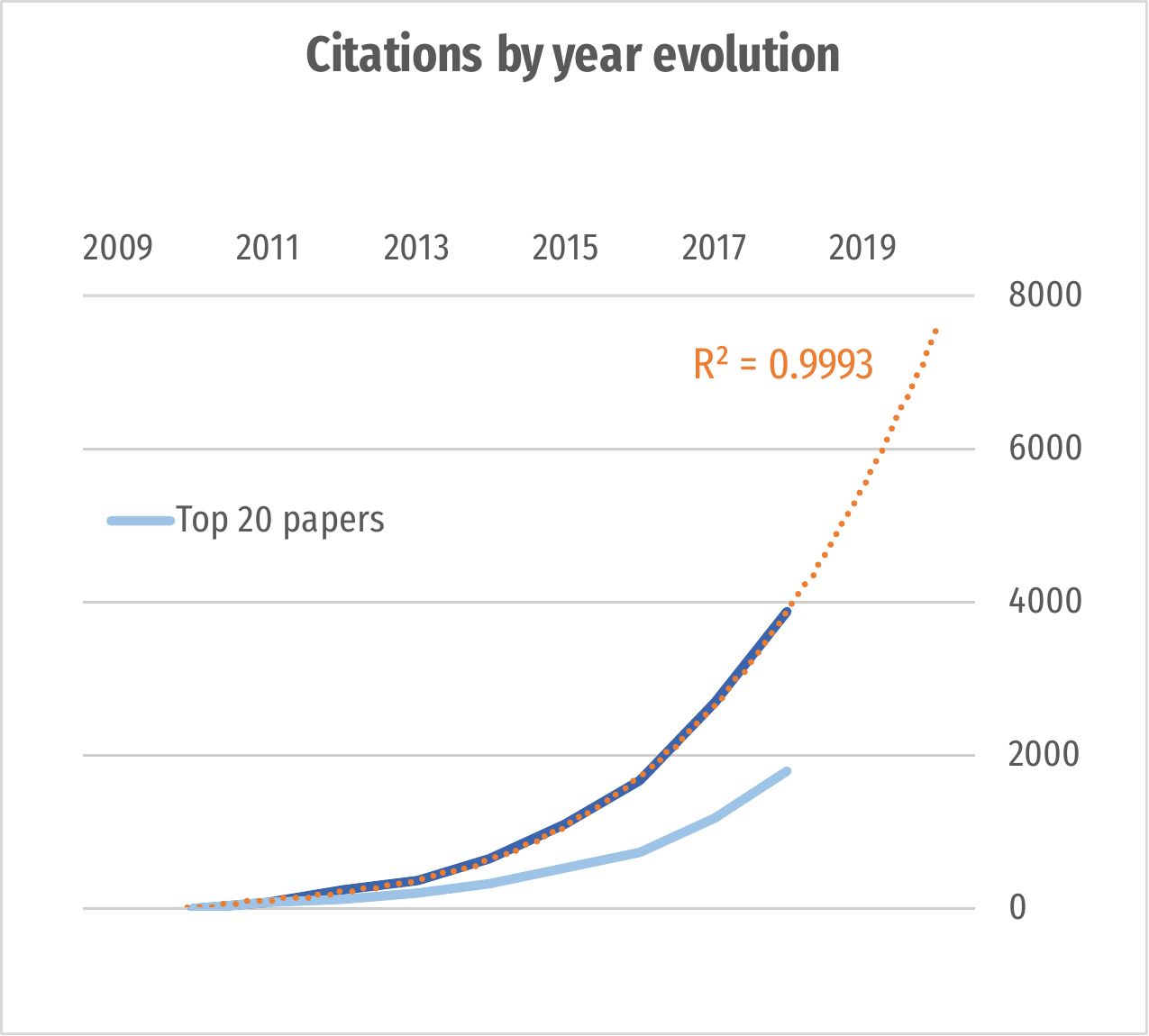} }}%
\caption{Evolution of research production (a), and citations (b), on \emph{Transfer Learning} in the last 10 years, and exponential growth projection with high coefficient of determination ($R^2$), despite publication decay in 2018. In (b) one can also see that the top 20 most cited papers are responsible for almost half the citations. (Data source: Web of Science, march/2019)}
\label{fig:citations_per_year}%
\end{figure}

This context reveals the huge unreached potential of Transfer Learning (TL), which aims to leverage prior experience to efficiently and effectively learn new tasks.
In practice, TL tends to be applied on an \textit{ad hoc} basis, where the transfer methods are simple extensions of the learning algorithms used \citep{torrey}. Such importance with a lack of consolidated methods and theories indicates this is a promising field for research. As Andrew Ng~\citep{ANg} says: ``Transfer Learning will be the next driver of Machine Learning success across industries''. From this perspective, it is understandable the growing interest on the subject (Figure \ref{fig:citations_per_year}).

\subsection{Objectives}\label{goals}
Our research questions are:
\begin{enumerate}[Q1.]
\item{What are the research frontiers in Transfer Learning?}
\item {Is it possible to base this evaluation on bibliometrics?}
\end{enumerate}
To answer these questions, we will first review the literature and reveal the main contributions to the field and how they relate to each other.

\subsection{Contributions}
\begin{enumerate}[C1.]
\item We present an updated systematic review of the literature on Transfer Learning, using the TEMAC framework (Section \ref{TEMAC}). A method that helps us focus on high impact contributions.
\item We extended TEMAC to analyze linguistic variations of abstracts using ScatterText ~\citep{kessler2017scattertext}, which, up to our knowledge, is an original usage of this visualization tool.
\item We identify, with bibliometric analysis support, the directions of research frontiers, and the open problems of the field. 
\end{enumerate}

\subsection{Overview}
In this brief introduction, we will present \emph{Related Works}. In the next section (section \ref{TEMAC}), we will explain our research method and the quantitative analysis to support our findings. In section \ref{literature}, the results, which are indeed the \emph{Literature Review}, are shown. \emph{Open Problems} are discussed in section \ref{open_problems}. Finally, we conclude in section \ref{conclusion}, presenting answers for our research questions.

\subsection{Related Work}\label{related}
As we specify in section \ref{literature}, there are already literature reviews in the field of \emph{Transfer Learning}, and they tend to be well cited. Noteworthily, \citet{PanYang}'s survey is the most cited paper in the field, and \cite{hohman2018visual} systematically reviews the literature on visual analytics in \emph{deep learning} with the gripping method of five W's and one H (Why, Who, What, How, When, and Where). Still, to the best of our knowledge, our literature review is the first in \emph{Transfer Learning} to be based on a bibliometric method. Similar to what we propose in our field, \cite{park2018identifying} applies a meta-analytical approach to identify research frontiers in \emph{Pattern Recognition}, and \cite{bhattacharya1997cross} uses bibliometrics to identify research frontiers in \emph{Physics}.

\section{Method: Literature review with a quantitative approach}\label{TEMAC}
Our literature review uses the bibliometric approach of ~\citet{mariano2017revisao} (TEMAC), which aims to give quantitative support to literature selection.

TEMAC consists of:
\begin{enumerate}[a)]
\item research preparation;
\item data presentation and interrelationships;
\item detailing, synthesis and validation.
\end{enumerate}

\subsection{Research Preparation}
\begin{figure}[htp]
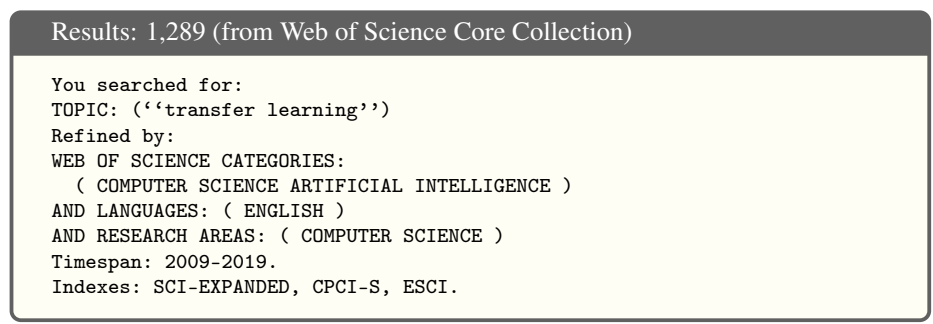


\begin{tcolorbox}[colback=yellow!5!white,colframe=gray!75!black,title={Results: 1,289 (from Web of Science Core Collection)}]
\footnotesize{
\begin{verbatim}
You searched for: 
TOPIC: (``transfer learning'')
Refined by: 
WEB OF SCIENCE CATEGORIES: 
  ( COMPUTER SCIENCE ARTIFICIAL INTELLIGENCE )
AND LANGUAGES: ( ENGLISH ) 
AND RESEARCH AREAS: ( COMPUTER SCIENCE )
Timespan: 2009-2019. 
Indexes: SCI-EXPANDED, CPCI-S, ESCI.
\end{verbatim}
}

\end{tcolorbox}
\caption{\emph{``10yearsSearch''}: Search parameters on \emph{Web of Science}.}
\label{card:wos}
\end{figure}

On \nth{31} of March, a search on \emph{Clarivate Analytics Web of Science (WoS)} shown in Figure  \ref{card:wos} resulted in 1,289 articles found. It is noticeable that the interest in the subject is thriving (Figure \ref{fig:citations_per_year}), and it is possible to project that in three years the number of articles will double (exponential growth, coefficient of determination: $R^2=0,9789$).

\begin{figure}[htp]
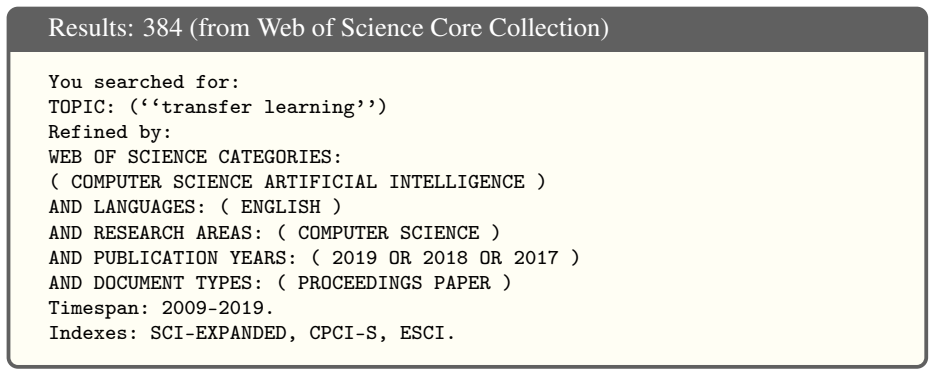

\begin{tcolorbox}[colback=yellow!5!white,colframe=gray!75!black,title={Results: 384 (from Web of Science Core Collection)}]
\footnotesize{
\begin{verbatim}
You searched for: 
TOPIC: (``transfer learning'')
Refined by: 
WEB OF SCIENCE CATEGORIES: 
( COMPUTER SCIENCE ARTIFICIAL INTELLIGENCE )
AND LANGUAGES: ( ENGLISH ) 
AND RESEARCH AREAS: ( COMPUTER SCIENCE )
AND PUBLICATION YEARS: ( 2019 OR 2018 OR 2017 )
AND DOCUMENT TYPES: ( PROCEEDINGS PAPER )
Timespan: 2009-2019. 
Indexes: SCI-EXPANDED, CPCI-S, ESCI.
\end{verbatim}
}

\end{tcolorbox}
\caption{\emph{``3yearsSearch''}: Search parameters for frontier analysis.}
\label{card:sota}
\end{figure}

A restricted search with only recent works (Figure \ref{card:sota}) was made for the section \ref{fronteiras} (\emph{Research Frontiers}).

\subsection{Data presentation and interrelationship}\label{inter}
\begin{figure*}[htp]
\centering
\fbox{\includegraphics[width=\textheight, angle=90]{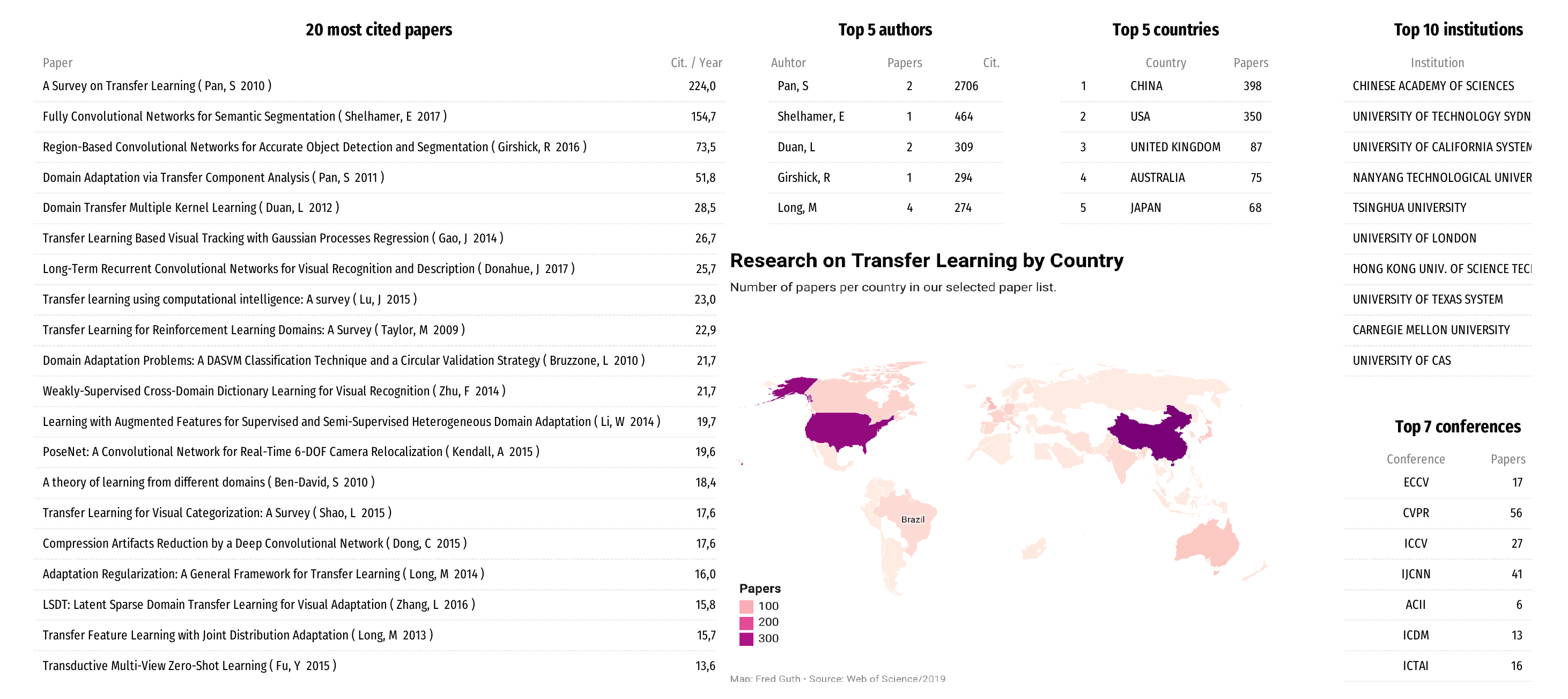}}
\caption{Overview of transfer learning research.}
\label{fig:toptop}
\end{figure*}

In this stage we analyse (see \S\ref{sec:panorama}):
\begin{enumerate}
\item most cited articles (Figure \ref{fig:toptop});
\item number of articles evolution year by year(Figure \ref{fig:citations_per_year}(a));
\item citations evolution year by year (Figure \ref{fig:citations_per_year}(b));
\item most published and cited authors (Figure \ref{fig:toptop});
\item most published and cited conferences (Figure \ref{fig:toptop});
\item most published and cited institutions (Figure \ref{fig:toptop});
\item countries by research production (Figure \ref{fig:toptop});
\item keyword frequency (Figures \ref{fig:clouds} and \ref{fig:scatterText}).
\end{enumerate}
\subsection{Synthesis and Validation}\label{detalha}

\begin{enumerate}[a)]
\item{\textbf{Co-citation analysis}:
Co-citation measures the frequency in which two papers are cited in the same reference list, and it is assumed that they are ``pieces'' of the same ``knowledge structure''. Co-citation analysis, therefore, maps the intellectual inheritance of a research field by identifying impactful works, but neglects the research frontier as those works had less time to be cited \citep{Vogel2012}.\\ The free software VOSviewer \citep{VOSviewer} was used to cluster works cited by the \emph{``10yearsSearch''} selected articles. With that, three knowledge clusters where identified (Figure \ref{fig:classicos}).}

\item{\textbf{bi-coupling analysis}:
Bibliographic coupling occurs when two papers have at least one reference in common. Papers are, thus, said to be coupled if their references overlap \citep{Vogel2012}. As it is possible to state a chronological order among cited and citing works, bibliographic coupling allows us to map research ``generations'' and, therefore, identify which research is in the leaves of this tree, i.e., are in the frontier of the field. It is important to note that, in this context, being in the frontier is just a chronological incident and does not mean it is a promising work. This limiting aspect of the quantitative approach points to the need for a qualitative complement to identify the ``classics of the future''. \\ In the TEMAC framework, the bibliographic coupling should be done in a period not longer than the last three years. In our analysis, we restricted the period from 2017 to March 2019 and the only selected works from conference proceedings (Figure \ref{card:sota}), assuming that these works have a shorter period of review and publication and, therefore, represent what is newer in the field. }
\item{\textbf{Textual analysis (\emph{bag-of-words}  with tf-idf)}:in this analysis articles are viewed as \emph{bag-of-words} and the concept of \textbf{tf-idf} is used to define which words better identify each paper. For instance, which words better explain research in frontier \emph{vis-à-vis} those which explain articles from \emph{``10yearsSearch''} (see Figure \ref{fig:scatterText}). The metric \textbf{tf-idf} is defined as:
\begin{equation}
\mathrm{tfidf}(t,d,D) = \mathrm{tf}(t,d) \cdot \mathrm{idf}(t, D)
\end{equation}
where $\mathrm{tf}(t,d)$ is the term frequency $t$ in document $d$ and $\mathrm{idf}(t, D)$ is the inverse of the frequency of $t$ in the document set $D$ (\emph{corpus}).
\begin{equation}
\mathrm{tf}(t,d) = 0.5 + 0.5 \cdot  \frac{f_{t, d}}{\max\{f_{t', d}:t' \in d\}}
\end{equation}
\begin{equation}
\mathrm{idf}(t, D) =  \log \frac{N}{|\{d \in D: t \in d\}|}
\end{equation}
where:
\begin{description}
\item $N$: size of corpus in number of documents $N = {|D|}$
\item$ |\{d \in D: t \in d\}| $ : number of documents where term  $ t $ appears (i.e., $ \mathrm{tf}(t,d) \neq 0$). To avoid division by zero if the term is not in the corpus, the denominator is adjusted to: $1 + |\{d \in D: t \in d\}|$.
\end{description}

The metric \textbf{tf-idf} is the basis of the visualization tool \emph{ScatterText}~\citep{kessler2017scattertext} which was used to generate the Figures \ref{fig:scatterText_documentbased} and \ref{fig:scatterText}.
}\label{analiseTextual}
\end{enumerate}

\section{Literature Review}\label{literature}
\subsection{A brief history of Transfer Learning}
Since 1995, when a NIPS\footnote{Currently, the Neural Information Process Systems conference is called NeurIPS.} workshop on ``Learning to Learn'' discussed the need for machine learning to retain and reuse previously acquired knowledge, research on transfer learning, although sometimes called by different names (\emph{learning to learn, life-long learning, knowledge transfer}) has attracted more and more attention~\citep{PanYang} (see Figure \ref{fig:citations_per_year}).

In 2005, a DARPA project announcement used, maybe for the first time, the term \emph{transfer learning}, defined as the goal of extracting knowledge from one or more \emph{source tasks} and applying it to \emph{target tasks}~\citep{PanYang}. A search on Web of Science can confirm that the first articles to use the term ``transfer learning'' appear in 2005.

In 2012, a deep neural network used by Alex Krizhevsky and team in the ImageNet Challenge (ILSVRC) was 41\% better than the second place, an outstanding result that ignited the exponential growth on deep learning research. Such success highlighted the importance of data availability for the advancement of artificial intelligence. It gave birth to a new era in transfer learning. Despite the cost of learning with big datasets like ImageNet, trained models proved to be easily suitable to initialize models for different tasks ({Ruder2019Neural, donahue2014decaf}). This ``fine-tunning'' approach allows good results on many tasks with orders of magnitude less data (see Section \ref{terceira_onda}).

In the present moment, \emph{transfer learning} is a customary topic in prestigious conferences like CVPR, ICCV, ICPR e NeurIPS (see Figure \ref{fig:toptop}, \emph{Top 10 conferences}).
\subsection{Notation and Definitions}
Transfer Learning entails the concepts of domain and task. According to the notation of \citet{PanYang}, a domain $\mathcal{D}$ is composed by a feature space $\mathcal{X}\subset R^d$ and a marginal distribution $P(\mathrm{X})$, from where samples $\mathrm{S}=\{x_1, \cdots, x_n\}\in\mathcal{X} $ are drawn. In an image classification problem, for instance, $\mathcal{X}$ is the space of all possible images with a certain dimension and number of channels, $x_i$ is an image, and $\mathrm{S}$ is the training \textit{dataset}.

Given a domain $\mathcal{D}=\{\mathcal{X}, P(X)\}$, a task $\mathcal{T}$ can be statistically defined by the conditional distribution $P(\mathrm{Y}|\mathrm{X})$, that is, $\mathcal{T}=\{\mathcal{Y}, f(\cdot)\}$, where $f(\cdot)$ is a goal function that given $x_i \in \mathcal{D}$, predicts its corresponding $y_i \in \mathcal{Y}$. 

Be $\mathcal{D}_S$ the source domain and $\mathcal{T}_S$ the source task, $\mathcal{D}_T$ the target domain and $\mathcal{T}_T$ the target task, \textbf{transfer learning } aims to help learning the function $f_T(\cdot)$ in $\mathcal{D}_T$ using knowledge from $\mathcal{D}_S$ and $\mathcal{T}_S$, where $\mathcal{D}_S\neq\mathcal{D}_T$ or $\mathcal{T}_S\neq\mathcal{T}_T$.

\subsection{An overview of transfer learning research}\label{sec:panorama}

Pan, S. is the most cited author (see Figure \ref{fig:toptop}, Top 10 authors) with 2706 citations. Such impact is mainly due to \emph{A Survey on Transfer Learning} \citep{PanYang}, which is the most cited article in the field with 2240 citations. The main contributions of this article are to present definitions, notations, and taxonomy for transfer learning that made sense for the research community. It was published in IEEE, a journal with an impact factor of 2,775, and that ranks at the 33rd position on InCite JCR for the \emph{Computer Science, Artificial Intelligence} category, which means it is not a usual publication for \emph{transfer learning} research.

China is the most productive country in the field, followed by the United States and the UK.

Most articles are published in conference proceedings, 63\%. CVPR\footnote{Conference on Computer Vision and Pattern Recognition} is the conference with most articles on \emph{Transfer Learning}, but ECCV\footnote{European Conference on Computer Vision}, despite ranking only 4\textsuperscript{th} in production, is the most cited conference. It is also notable that computer vision conferences have been the most popular venue for TL research. It is worth mentioning the lack of conferences with NLP focus in this list and that among the 20 most cited articles, none is on language.

\subsection{The classics}\label{classicos}
\begin{figure}[h]
\fbox{\includegraphics[width=\columnwidth]{completo_21.pdf}}
\source{Web of Science (march/2019)}{VosViewer\protect{(~\cite{VOSviewer}}}
\caption{Knowledge clusters from the co-citation analysis. Clusters represent papers that are normally cited together in the list of references of the 10yearsSearch articles.}
\label{fig:classicos}
\end{figure}

Co-citation analysis using \emph{VosViewer} shows 3 knowledge clusters. They present a strong temporal component and can be seen as waves: the first encompasses publications prior to 2011 (the mode is 2006), shown in yellow in Figure \ref{fig:classicos}; the second goes from 2011 to 2014 (mode 2012) and is in red; and the third and last wave, green in the figure, includes articles from 2012 to present day (mode 2014).
\subsubsection{First wave}\label{firstwave}
One of the main characteristics of this wave is the strength of its theoretical component. Some of its works present whole classes of transfer learning: \citet{thrun1996learning} and \citet{Caruana1997} introduce \emph{Multi-Task Learning} and the idea that auxiliary tasks introduce inductive bias that helps the learning convergence; \citet{Chapelle:2010}'s book on \emph{Semi-Supervised Learning}; \citet{Raina2007} on \emph{Self-taught learning}; \citet{Vapnik1998} and the fundamentals of statistical learning theory.

Another component of this wave is the Domain Adaptation articles that aim to learn latent characteristics of reduced dimensionality, assuming that in the latent space source and target domains are similar: \citet{ando2005framework}, \citet{Blitzer:2006:DAS:1610075.1610094} and \citet{DaumeIII2006}, for instance.

Lastly, some of the most cited articles of this wave are surveys: \citet{PanYang}, \citet{Taylor:2009:TLR:1577069.1755839}; not by coincidence they were published by the end of the wave, organizing the research material up to that point in time.

\subsubsection{Second wave}\label{second_wave}
Articles on Domain Adaptation dominated the second wave. Some kept the focus on latent characteristics like \citet{Pan2011}, but most articles try to select samples from source domain that are similar to the target domain: 
\citet{BenDavid2009} approach distribution similarities of domains distributions from a theoretical standpoint; \citet{SiSi2010} tries to minimize divergence among domains; Common to this trend of work are approaches based on sample classifiers, mostly \emph{Support Vector Machines} (SVMs): \citet{Yang2007}, \citet{Bruzzone2010}, \citet{LixinDuan2012}, among others.

We also find works on unsupervised domain adaptation: \citet{BoqingGong2012}, \citet{Fernando2013}.

It is noteworthy that among the articles in our search, there was no \emph{survey} ``closing'' this wave. An excellent example of this kind of work would be the chapter \emph{A Comprehensive Survey on Domain Adaptation for Visual Applications} \citep{Csurka2017}. Unfortunately, this work is not indexed by \emph{WoS} and, therefore, could not be found in our search results.

\subsubsection{Third wave}\label{terceira_onda}
The third wave cluster includes transfer learning approaches in \emph{deep learning} context, which we can call \emph{deep transfer learning}.

Here we find classical deep learning literature like: \citet{Hinton2006}, maybe the seminal article on \emph{Deep Learning}; \citet{Bengio2009} presents \emph{representation learning} and curriculum, a list of tasks to be learned in sequence, because they have growing levels of complexity; \citet{LeCun2015}'s review on \emph{Deep Learning} for \emph{Nature}. Here we also find \citet{alexnet}'s article on \emph{AlexNet}, which by winning the 2012 ImageNet challenge (ILSVRC) with a margin of almost 40\% over the second place, gave birth to the current ``gold rush'' of \emph{Deep Learning}.

Some articles in this wave are about large datasets, an essential component of deep learning success: 

\citet{Deng2009} presents ImageNet and suggest that models learned on ImageNet can be used to more efficiently learn new domains; 

\citet{Everingham2009} is about Pascal VOC; \citet{Russakovsky2015} analyses ImageNet's impact on different computer vision problems and, therefore, the role of transfer learning.

There are also articles that present model architectures that use pre-trained models for feature extraction or can be used after training as such: \citet{Girshick2014}, which presents the R-CNN model, \citet{He2016}, ResNet, and \citet{simonyan2014very}, object detection; \citet{JialueFan2010} human tracking; \citet{Long2015} on semantic segmentation; among others.

Transfer Learning became so ubiquitous in \emph{Deep Learning} that \citet{mahajan2018exploring} claims that \emph{not} pre-training models with ImageNet in computer vision problems is now considered foolhardy.

For this reason, it is quite difficult to point out articles with the sole focus on transfer learning. Some worth mentioning are: \citet{glorot2011domain} which proposes using \emph{deep neural networks} to learn common representation among domains; \citet{donahue2014decaf} and \citet{Oquab:2014:LTM:2679600.2680210} which give some theoretical support to the fine-tuning approach.

\subsubsection{Textual analysis results}
The 20 most cited articles represent almost half the citations (Figure \ref{fig:citations_per_year}b). Consequently, we assume that a textual analysis of this subset of \emph{``10yearsSearch''} is a good proxy for the whole.

\begin{figure}
\fbox{\includegraphics[width=\columnwidth]{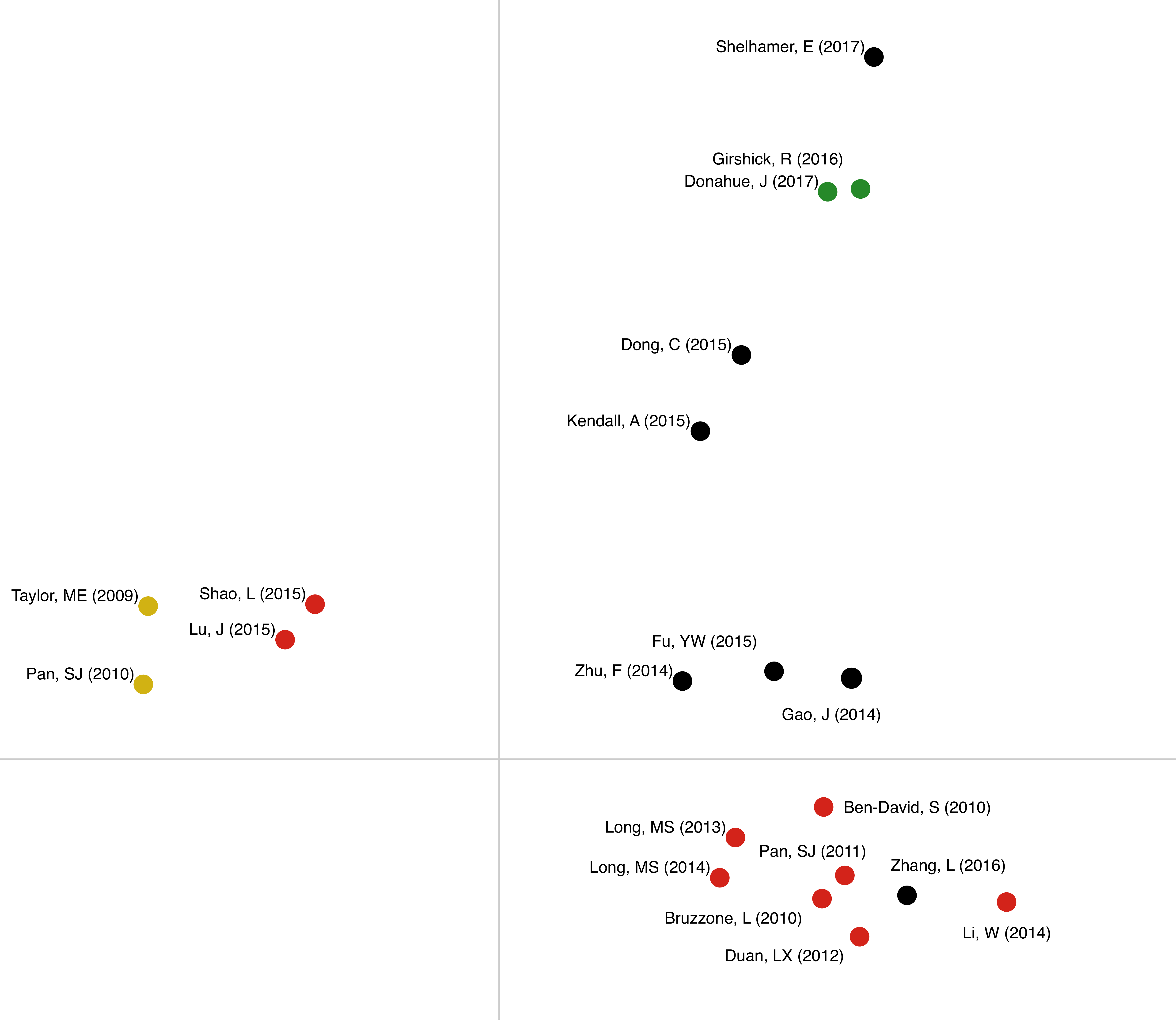}}
\source{WoS (março/2019)}{ScatterText~\protect({\cite{kessler2017scattertext}}}
\caption{Top 20 most cited articles visualized by proximity of \emph{bag of words}.} \label{fig:scatterText_documentbased}
\end{figure}

By a different method, we approximately cluster in the same way as before. In the Figure \ref{fig:scatterText_documentbased}, each quadrant represents a cluster in Figure \ref{fig:classicos}. That is a strong validation of this analysis.

\begin{figure}
\fbox{\includegraphics[width=\columnwidth]{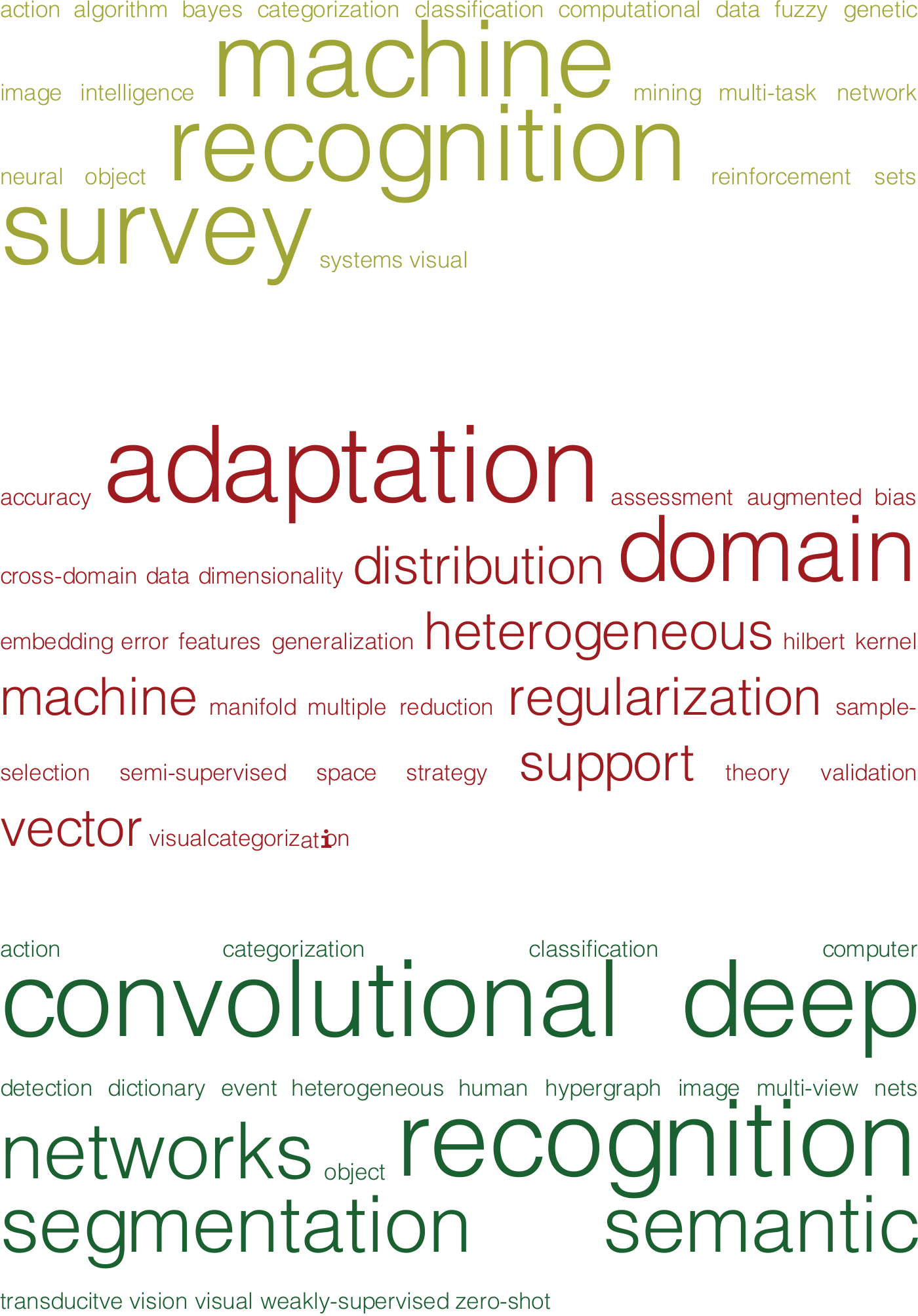}}
\source{WoS (march/2019)}{TagCrowd}
\caption{Word clouds for all quadrants of chart \ref{fig:scatterText_documentbased}.} \label{fig:clouds}
\end{figure}

In Figure \ref{fig:clouds} it is possible to see the ``word cloud'' for each quadrant of chart in Figure \ref{fig:scatterText_documentbased}. Those terms correspond well with the qualitative analysis of Section \S\ref{classicos}.


\subsection{Research frontiers }\label{fronteiras}
To identify the research frontiers in transfer learning, two analysis were performed:
\subsubsection{Textual Analysis}

\begin{figure}
\includegraphics[width=\columnwidth]{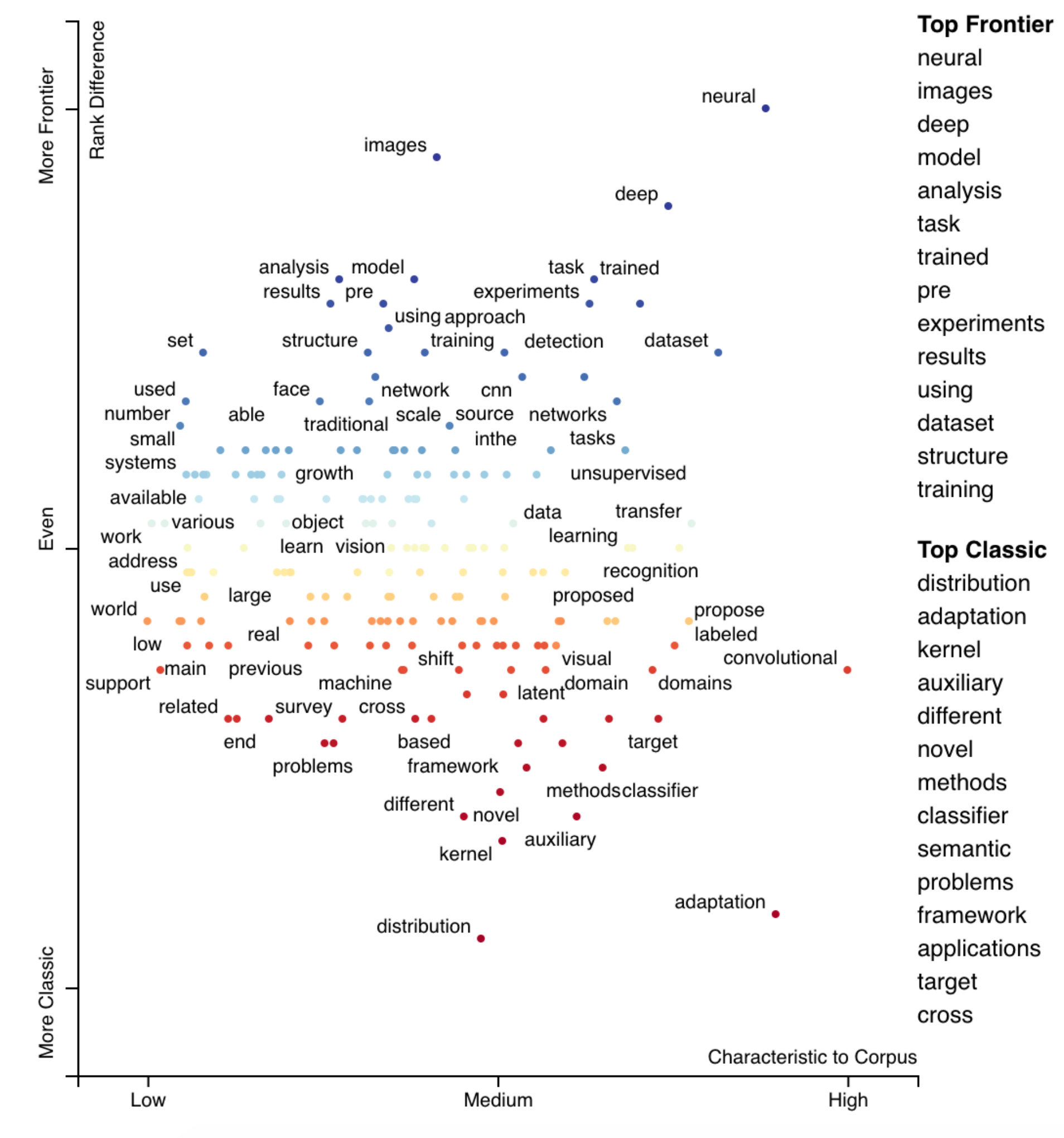}
\source{\emph{WoS}(march/2019)}{ScatterText\protect({\cite{kessler2017scattertext}}}
\caption{Visual analysis of ``frontier'' terms versus ``classic'' terms in the \emph{Transfer Learning} context.} \label{fig:scatterText}
\end{figure}

We used \emph{ScatterText}~\citep{kessler2017scattertext} to visualize the terms which better represent \emph{``3yearsSearch''} versus the ones that better represent \emph{``10yearsSearch''}, resulting in Figure \ref{fig:scatterText}.

The terms more related with the research frontier are \emph{deep}, \emph{neural}, \emph{images} and words like \emph{cnn}, \emph{trained networks} and \emph{datasets}. That matches our analysis of the third wave in \S\ref{terceira_onda}. The terms more related with \emph{``10yearsSearch''} are: \emph{distribution}, \emph{domain}, \emph{adaptation}, \emph{auxiliary} (from auxiliary tasks), \emph{kernel}; which are quite similar to what we found in second wave \S\ref{second_wave}.

\subsubsection{Bibliographical Coupling}
\begin{figure}[hb]
\fbox{\includegraphics[width=\columnwidth]{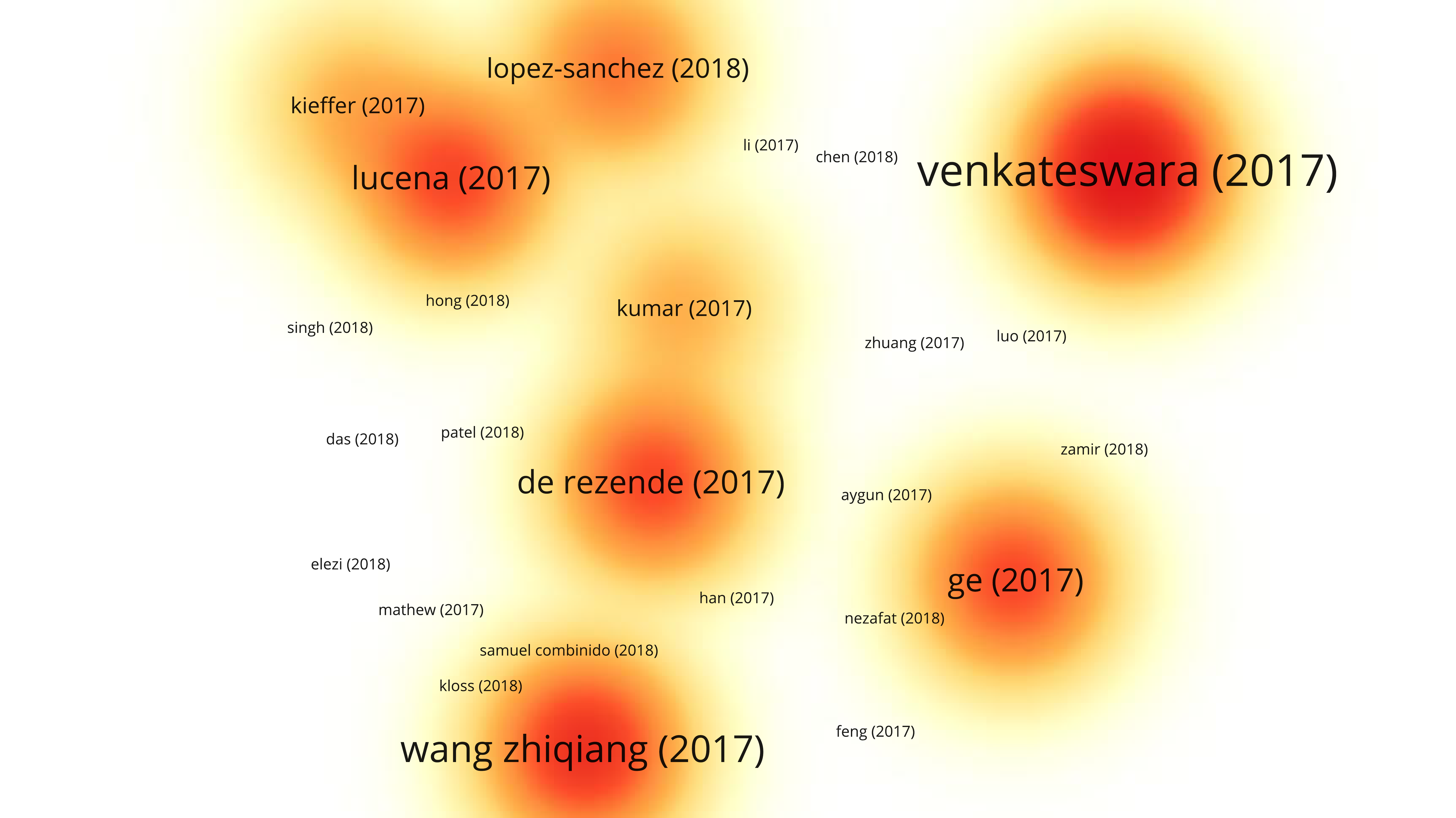}}
\source{\emph{WoS}(march/2019)}{VosViewer}
\caption{Heatmap of the bibliographical coupling analysis.} \label{fig:bicoupling}
\end{figure}

The result of the bibliographical coupling analysis using VosViewer \citep{VOSviewer} with \emph{''3yearsSearch''} articles can be seen in Figure \ref{fig:bicoupling}.

Among the clusters, we have:
\citet{Lucena2017} and \citet{Rezende2017}, both applications of \emph{deep convolutional networks} for new problems, \emph{face anti-spoofing} and \emph{fake images detection}, respectively.
\citet{Venkateswara2017} and \citet{Ge2017} propose new Domain Adaptation methods using \emph{Deep Learning}. Clearly every work in this frontier is somehow about \emph{Deep Transfer Learning}.

\subsubsection{The classics of tomorrow}
As stated previously, the quantitative analysis using bibliographical coupling can show which articles are in the frontier (the ones on Deep Transfer Learning) but not which of them are promising. For such, a qualitative analysis of candidate articles is crucial.
 
In this context, we here highlight some works that we believe have the potential to become classics of tomorrow:
\begin{itemize}
\item \citet{taskonomy}: one of the best papers of CVPR 2018, this paper explores relations between tasks, measuring them, and building a graph that can be used to define sequences of training that lead to smaller sample complexity for a specific task.
\item \citet{ulmfit}: by exploiting transfer learning, the ULMFit model was able to perform between 18 and 24\% better than the previous state of the art for some NLP problems and can become to language what \citet{alexnet} was for computer vision.
\item \citet{CycleGan}: propose \emph{Generative Adversarial Networks (GANs)} for unsupervised domain transfer. We are extremely optimistic regarding the potential of GANs in transfer learning.
\item \citet{Ruder2019Neural}: presents a new taxonomy for transfer learning in the context of NLP.
\end{itemize}

\section{Open Problems}\label{open_problems}
The overview of a systematic review allows us to perceive some gaps on the field accumulated knowledge. Some open problems are:
 \begin{enumerate}
 \item Metrics: no specific metrics for transfer learning.
 \item Taxonomy: current taxonomy \citep{PanYang} focuses too much on domain adaptation and too little on inductive transfer learning. Also, we need to include more novel ideas like GANs and \emph{autoencoders}.
\item NLP: there is still too little about transfer learning for NLP, with the exception made for \citet{Ruder2019Neural}.

\item Theory: on the first wave, the role of theory was trying to indicate promising approaches. Today, some approaches work well in practice, but the theory does not explain why. It is important to know why. 
\end{enumerate}

\section{Conclusion}\label{conclusion}
In this systematic literature review, it was shown that \emph{deep transfer learning} is in the research frontier of transfer learning. It is a very broad sub-field and encompasses a) usage of pre-trained models as feature extraction for new applications; b) finding latent representations among domains; c) unsupervised style transfer between domains; among others.
In this work, we were guided by the TEMAC method and extended it with other analytical tools to validate our conclusions. By doing so, it was possible to demonstrate that it is possible to identify research frontiers with a quantitative-based analysis on bibliographical data, which answers our research questions.

In future work, some important matters need to be addressed: First, \citet{PanYang} taxonomy is outdated: for obvious reasons, it focuses on the kind of work that had been done up to a decade ago and does not give much guidance for classifying articles in the research frontier. Besides, the systematic review method deserves some improvement:
1) It could include other bases as \emph{Scopus} and \emph{Google Scholar}; 
2) We could expand the research queries to include other terms: 
\emph{multi-task learning}, \emph{domain adaptation}, and even some that are not as used anymore like \emph{learning to learn} and \emph{lifelong learning}; 3) it could encompass the summarization of most important articles, preferably adopting a framework like the five W's and one H~\citep{hohman2018visual}.

Lastly, from this review it became clear to us the need to improve the theory of Deep Learning to better explain how and why the transfer of knowledge happens in deep neural networks.
%


%
%

\bibliographystyle{spbasic}      
\bibliography{references}

\end{document}